\documentclass[conference]{IEEEtran}
\IEEEoverridecommandlockouts

\usepackage{cite}
\usepackage{amsmath,amssymb,amsfonts}
\usepackage{algorithmic}
\usepackage{graphicx}
\usepackage{textcomp}
\usepackage{xcolor}
\usepackage[T1]{fontenc} 
\usepackage{array,color}
\usepackage[caption=false,font=normalsize,labelfont=sf,textfont=sf]{subfig}
\usepackage{stfloats}
\usepackage{url}
\usepackage{verbatim}
 \usepackage{bm}
 \usepackage{mathrsfs} 
\usepackage{times}
\def\BibTeX{{\rm B\kern-.05em{\sc i\kern-.025em b}\kern-.08em
    T\kern-.1667em\lower.7ex\hbox{E}\kern-.125emX}}
 \usepackage[letterpaper, top=1.92cm, bottom=4.2cm, left=1.52cm, right=1.52cm]{geometry}
\begin{document}

\title{Fast Time-Varying mmWave Channel Estimation: A Rank-Aware Matrix Completion Approach\\
\thanks{This work was funded by the National Nature Science Foundation of China under Grant 62271041 and in part by the Hongguoyuan Science and Technology Project under Grants WX-2025-0803, WX-2024-0718, and WX-2024-0805.}
}
\author{
\IEEEauthorblockN{Tianyu Jiang\IEEEauthorrefmark{1},  Yan Yang\IEEEauthorrefmark{1}, Hongjin 
Liu\IEEEauthorrefmark{2}, Runyu Han\IEEEauthorrefmark{1}, Bo Ai\IEEEauthorrefmark{1}, and Mohsen Guizani\IEEEauthorrefmark{3}}   
\IEEEauthorblockA{\IEEEauthorrefmark{1}School of Electronic and Information Engineering, Beijing Jiaotong University, Beijing, China} 
\IEEEauthorblockA{\IEEEauthorrefmark{1}\textit{jiangty, yyang, runyuhan, boai@bjtu.edu.cn}}
\IEEEauthorblockA{\IEEEauthorrefmark{2}Beijing SunWise Space Technology Ltd., SunWise Space, Beijing, China} 
\IEEEauthorblockA{\IEEEauthorrefmark{2}\textit{lhjbuaa@163.com}}
\IEEEauthorblockA{\IEEEauthorrefmark{3}Mohamed bin Zayed University of Artificial Intelligence Abu Dhabi, United Arab Emirates} 
\IEEEauthorblockA{\IEEEauthorrefmark{2}\textit{mguizani@ieee.org}}
}
\maketitle

\begin{abstract}
We consider the problem of high-dimensional channel estimation in fast time-varying millimeter-wave MIMO systems with a hybrid architecture. By exploiting the low-rank and sparsity properties of the channel matrix, we propose a two-phase compressed sensing framework consisting of observation matrix completion and channel matrix sparse recovery, respectively. First, we formulate the observation matrix completion problem as a low-rank matrix completion (LRMC) problem and develop a robust rank-one matrix completion (R1MC) algorithm that enables the matrix and its rank to iteratively update. This approach achieves high-precision completion of the observation matrix and explicit rank estimation without prior knowledge. Second, we devise a rank-aware batch orthogonal matching pursuit (OMP) method for achieving low-latency sparse channel recovery. To handle abrupt rank changes caused by user mobility, we establish a discrete-time autoregressive (AR) model that leverages the temporal rank correlation between continuous-time instances to obtain a complete observation matrix capable of perceiving rank changes for more accurate channel estimates. Simulation results confirm the effectiveness of the proposed channel estimation frame and demonstrate that our algorithms achieve state-of-the-art performance in low-rank matrix recovery with theoretical guarantees.
\end{abstract}

\section{Introduction}
\IEEEPARstart{B}{y} leveraging abundant spectrum resources, millimeter wave (mmWave) communications emerge as a key enabler for future wireless networks to meet ever-increasing capacity demands \cite{b1}. To combat the high propagation losses experienced in mmWave bands, a hybrid analog-digital (HAD) beamforming architecture is a low-complexity solution that can offer a high-gain directional beam, substantially addressing this radio propagation limitation. In mobile mmWave communication scenarios, the transmitter needs to form a beam that adapts to the receiver's movement so that the beam can maintain continuous beam alignment in the receiver's direction to ensure reliable transmission. To do this, the acquisition of precise channel estimates is essential. However, despite notable theoretical advancements, practical mmWave MIMO communication systems inherently involve high-dimensional channel matrices. In the case of high mobility, a complete measurement of channel state information (CSI) is often not practical; real-time measurement data is often outdated, even corrupted with large errors. Thus, accurate channel estimation remains difficult, especially in rapid channel variations under high mobility, such as the sixth-generation (6G) communication systems for vehicle-to-everything (V2X) communications, where terminals move at a speed of 120 km/h or even higher. 
\\
\indent
An alternative solution is to exploit the inherent sparsity of mmWave MIMO channels. So far, this property has been exploited in the literature and identified as low-rank matrix recovery/completion methods in compressed sensing [2-5]. The basic idea is to formulate the problem as a matrix rank minimization problem and solve it efficiently by nuclear-norm minimization. Theoretically, a low-rank channel matrix can be stably recovered from incomplete, inaccurate, and noisy observations by solving a rank minimization problem, if the observation matrix satisfies a restricted isometry property (RIP) [2]. Although classical orthogonal matching pursuit (OMP) is effective against the absence of small Gaussian noise in the data, it suffers from a high computational cost in sparse recovery that may render it impractical for high-dimensional mmWave channel matrices. Various adaptive techniques have been proposed to address the challenges posed by time-varying mmWave channel estimation. Specifically,  sparse Bayesian learning (SBL) based channel estimation in mmWave hybrid MIMO OFDM systems was proposed in \cite{b62} and employed to track the propagation paths. It is shown in \cite{b36} that the authors proposed a novel SBL-based group-sparse paradigm for mmWave MIMO OFDM channel estimation, and developed a low-complexity version to reduce the computational cost of the high-dimensional channel matrix. Most recently, this method could be viewed as a way to sense and recover a low-rank channel matrix [7]. Nevertheless, existing methods predominantly focus on static sparsity patterns, leaving out the effect of rank uncertainty in high-mobility scenarios \cite{b45,b43}.
\\
\indent
In brief, the contributions of this paper are as follows:
\begin{itemize}
\item By incorporating an $\ell_1$-norm regularization, we devise a robust rank-one matrix completion (R1MC) algorithm to achieve high-precision observation matrix completion and explicit rank estimation from incomplete or corrupted observations. 
\item We propose a novel compressed sensing (CS) framework for high-dimensional mmWave MIMO channel estimation. The channel reconstruction is accordingly formulated as a sparse matrix recovery problem. 
\item To achieve accurate channel estimation, we develop an adaptive rank estimation framework integrated with a discrete-time autoregressive (AR) model to capture the channel dynamics, enabling temporal tracking of rank variations without prior knowledge. 
\end{itemize}

The rest of the paper is organized as follows: Section II introduces the system model, the problem formulation, and a sparse representation of the channel in the angular domain. In Section III, a two-phase CS channel estimator is proposed.  Simulation results are presented in Section IV, and the conclusion is provided in Section V.
\section{System Model}

\begin{figure*}[t]
	\centering 
	\includegraphics[width=14.5cm]{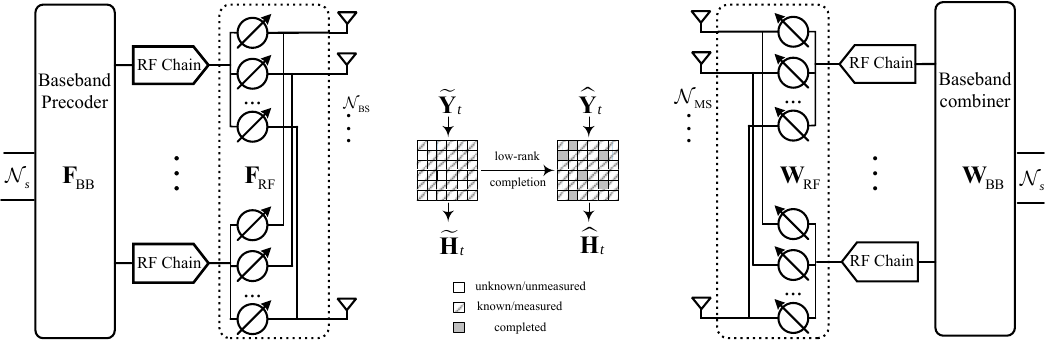}
	\caption{mmWave signal transmission model and time-varying channel estimation from incomplete observations.}
\end{figure*}
\subsection{Channel Model}
We consider a single-user mmWave MIMO system comprising a base station (BS) and a mobile station (MS). The BS employs ${{\mathcal N}_\text{BS}}$ antennas with ${{\mathcal M}_\text{BS}}$ RF chains, while the MS is equipped with ${{\mathcal N}_{\text{MS}}}$ antennas and ${{\mathcal M}_\text{MS}}$ RF chains. Both BS and MS adopt a hybrid architecture. Fig. 1 illustrates the end-to-end transmission structure, where $\widetilde{\textbf Y}_t$ and $\widehat{\textbf Y}_t$ represent incomplete- and complete observation matrix at time instance $t$, corresponding to coarse $\widetilde{\textbf H}_t$ and refined $\widehat{\textbf H}_t$, respectively. With a geometric wideband mmWave channel model, the beam space MIMO representation of the delay-$d$ MIMO channel matrix, ${{\textbf{H}}}(d)$, can be written as [6]
\\
\indent
%
\begin{align}{{\textbf H}}[d]=\,&\sqrt{\frac{{\mathcal N}_{{\text {BS}}}{\mathcal N}_{\text {MS}}}{{\rm L}_{\rm P}}}\sum_{\ell=1}^{\mathcal L} \sum_{\kappa_\ell=1}^{{\mathcal K}_\ell} \alpha_{\kappa_\ell} \varrho\left(d {\mathcal T}_{\rm s}-\tau_\ell -\tau_{\kappa_\ell}\right) \notag \\&\times {\textbf a}_{\text {MS}}\left(\theta_\ell - \mathcal{\vartheta}_{\kappa_\ell}\right) {\textbf a}^\dagger_{\text {BS}} \left(\phi_\ell - \varphi_{\kappa_\ell} \right),\end{align}
where $\mathcal L$ is the number of paths, $\alpha_{\kappa_\ell}$ is the complex gain associated with the $\ell$th cluster, and each cluster is assumed to contribute $\kappa_{\ell}$ paths between the BS and the MS; $\varrho(\cdot)$ is a pulse-shaping filter for $\mathcal T_{\text s}$-spaced signaling generated at $d {\mathcal T}_{\rm s}-\tau_\ell -\tau_{\kappa_\ell}$ seconds; ${\theta }_{\ell}\in [0,\pi]$ and ${\phi }_{\ell}\in [0,\pi]$ are the associated azimuth AoAs and azimuth AoDs respectively; $\vartheta_{{\kappa_\ell}}$ denotes the offset of the $\kappa$th path compared to the mean AoDs with the $\ell$th cluster; $\varphi_{\kappa_{\ell}}$ denotes the offset of the $\kappa$th path compared to the mean AoAs with the $\ell$th cluster; Superscripts $(\cdot)^\dagger$ stands for the Hermitian transpose. 
Suppose a uniform linear array (ULA) antenna is used, the beamforming gain at the BS and MS is respectively given by
\begin{equation}
\begin{aligned}
& {{\textbf{a}}_{\text{MS}}}({{\theta_\ell }})=\frac{1}{\sqrt{{{\mathcal N}_\text{MS}}}}{[1,{{e}^{j\frac{2\pi }{\lambda_c }\delta\sin {{\theta_\ell }}}},\ldots ,{{e}^{j({{\mathcal N}_{\text{MS}}-1)\frac{2\pi }{{\lambda_c} }\delta\sin {{\theta_\ell }}}}]}^{{T}}}\\ 
& {{\textbf{a}}_{\text{BS}}}({{\phi_\ell }})=\frac{1}{\sqrt{{{\mathcal N}_\text{BS}}}}{{[1,{{e}^{j\frac{2\pi }{\lambda_c }\delta \sin {{\phi_\ell }}}},\ldots ,{{e}^{j({{\mathcal N}_\text{BS}}-1)\frac{2\pi }{\lambda_c}\delta \sin {{\phi_\ell }}}}]}^{{T}}},
\end{aligned}
\end{equation}
where ${{\textbf{a}}_{\text{MS}}}$ and ${{\textbf{a}}_{\text{BS}}}$ denote the antenna array response vectors of the MS and BS, respectively. Here, Superscripts $(\cdot)^T$ stands for the transpose; $\lambda_c$ is the signal wavelength, and $\delta$ is the distance between neighboring antenna elements, which is usually set to $\delta=\lambda_c/2$.  As a result, the delay-$d$ channel model in (1) can be represented as
\begin{equation}{\textbf H}=\sum_{d=0}^{D-1} \textbf{{H}}[d] e^{-j 2 \pi d}.\end{equation}
This clustered channel model exploits the fact that each cluster comprises multiple paths. It leaves out the low-rank structure inherently induced by the clustering of paths. It verifies that the considered time-varying channel matrix \(\textbf{H}_t\) has a rank equivalent to its sparsity level, assuming the grid mismatch issue is disregarded. Thus, if the channel matrix rank can be precisely known, a priori knowledge of the sparsity can be provided for the OMP algorithm.
\\
\indent
We further process the signal model by first exploiting the sparsity property in the angular domain of the channel, which can better deal with the compressed sensing (CS) problem in the time domain. For the objective of sparse recovery,  the matrices ${\textbf{A}}_{\text{BS}} \in \mathbb{R}^{\mathcal N_\text{BS} \times \mathcal L}$ and ${\textbf{A}}_{\text{MS}} \in \mathbb{R}^{\mathcal N_\text{MS} \times \mathcal L}$ are introduced, which can be pre-computed at the MS. At time instant $t$, the channel model presented in (3) can be reformulated in a more concise form:
\begin{equation}
	{{\textbf{H}}_t}={{\textbf{A}}_{\text{MS}}}{\overline{\textbf{H}}_t}\textbf{A}_{\text{BS}}^{\dagger}\quad \forall\, t=1,2,...,T,
\end{equation}
where ${{\textbf{A}}_{\text{MS}}}=\{ {{\textbf{a}}_{\text{MS}}}({{\phi_\ell }})\}_{\ell=1}^{\mathcal L_1}$ and ${{\textbf{A}}_{\text{BS}}}=\{ {{\textbf{a}}_{\text{BS}}}({{\theta_\ell }})\}_{\ell=1}^{\mathcal L_2}$ is an overcomplete matrix; ${\overline{\textbf{H}}_t}\in \mathbb C^{\mathcal L_1\times \mathcal L_2}$ is a sparse matrix, i.e., a sum of $\mathcal L$ sparse matrices.
\subsection{Signal Transmission Model}
The paper focuses on the estimation of time-varying channels in the angular domain. For the $t$th time instances, the BS applies an RF chain $\textbf{F}_t=\textbf{F}_{\text{RF}} \textbf{F}_{\text{BB}}\in \mathbb{C}^{{\mathcal N}_\text{BS} \times {\mathcal M}_\text{BS}}$, which can be implemented by quantizing the angles at the analog phase shifters. The transmitted signal for the $t$th time instance in the absence of noise is
\begin{equation}
\textbf{x}_t=\textbf{F}_t\textbf{s}_t, 
\end{equation}
where $\textbf{s}_t \in \mathbb{C}^{{\mathcal N}_s\times 1}$ is the symbol vector at time instances $t$; $\mathcal N_s$ represents the number of transmittable data streams. 
\\
\indent
From (1), the unknown channel parameters within $\textbf{H}_t$ are highly interdependent due to multi-path fading and can even be suppressed due to non-coherent summation. However, we still process the signal in a better presentation and thus sparse recovery by integrating the frame-structured channel and the specific measurement matrix in the following. During the $t$th time instances, the MS uses an RF chain ${{\textbf{W}_t}}={{\textbf{W}}_{\text{RF}}}{{\textbf{W}}_{\text{BB}}}\in {{\mathbb{C}}^{{{\mathcal N}_\text{MS}}\times {\mathcal M_\text{MS}}}}$, which is implemented using quantized phase shifts at the receiver side. In this way, the resulting post-combining signal is given by
\begin{equation}\textbf{y}_t=\textbf{W}^\dagger_t{\textbf{H}}_t{\textbf{F}}_t\textbf{s}_t+\textbf{n}_t.\end{equation}
where $\textbf{n}_t$ denoted additive noise vector obeying Gaussian distribution $\mathcal{N}(0, {\sigma }^{2}\textbf{I})$. Once $M$ training symbol vectors at time instance $t$ are completed, i.e., ${\textbf s}_t=[{\textbf s}^T_t[1],{\textbf s}^T_t[2],\dots,{\textbf s}^T_t[M]]^T$, ${\textbf y}_t=[{\textbf y}^T_t[1],{\textbf y}^T_t[2],\dots,{\textbf y}^T_t[M]]^T$, the training and observation matrix  is respectively defined as
\begin{equation} 
\begin{aligned}
{\textbf {S}}_t&=\left [{\textbf {s}}_t[:,1], {\textbf {s}}_t[:,2], \dots, {\textbf {s}}_{t}[:,\mathcal N_{\text {BS}}]\right ],\\
{\textbf {Y}}_t&=\left [{\textbf {y}}_t[:,1], {\textbf {y}}_t[:,2], \dots, {\textbf {y}}_{t}[:,\mathcal N_{\text {BS}}]\right ].
\end{aligned}
\end{equation}
\indent
In this paper, the precoder and combiner structures are configured to be high-resolution phase shifters (PSs) to realize the analog beamformers. 
For the sake of notation, we denote ${\textbf F}_t^T\otimes {\textbf W}^{\dagger}_t$ by ${\bm\Phi}_t$, such that (6) can be rewritten as:
\begin{equation}
\begin{aligned}
{{{\textbf{Y}}_t}}&={\textbf W}^{\dagger}_t{{\textbf H}}_t{\textbf F}_t{{\textbf{S}_t}}+{{\textbf{N}_t}}\\
&=\left ({{\textbf F}_t{\textbf{S}_t}}\right )^T\otimes {\textbf W}^{\dagger}_t{{\textbf H}}_t+{{\textbf{N}_t}}\\
&\triangleq {\bm \Phi}_t{{\textbf H}}_t+{{\textbf{N}_t}},
\end{aligned}
\end{equation}
where $\otimes$ represents Kronecker product, $\bm{\Phi}_t$ denotes the measurement matrix, i.e., constructed dictionary, ${\textbf{Y}_t}$ and ${\textbf{H}_t}$ are reorganized into matrices according to the original dimensions\cite{b433}.  Considering a normalized symbol set $\vert\vert {\textbf S}_t\vert\vert=\textbf I$, we have $\bm{\Phi}_t={\textbf F}_t^{T}\otimes {\textbf W}_t^{\dagger}$. The observation matrix considered in this paper is incomplete or corrupted. In this sense, the channel estimation from incomplete observations becomes a matrix completion problem, i.e., a sparse signal recovery problem. 
\subsection{Problem Formulation}
For mmWave bands, the number of paths $\mathcal L$ is usually small. The observation model in (4) can thus be represented as a sampling process from a low-rank matrix.  The incomplete observation can be expressed as sampling from a low-rank matrix [2]:
\begin{equation}\label{LRMC}
\begin{aligned}
(\widetilde{\textbf Y}_t)_{ij}=({\textbf W}_t^{\dagger}{\textbf H}_t{\textbf F}_t)_{ij}\quad (i,j)\in \bm\Omega, 
\end{aligned}
\end{equation}
where $(\widetilde{\textbf Y}_t)_{ij}$ denotes the $(i, j )$th entry of $\widetilde{\textbf Y}_t$, and $\bm\Omega$ denotes a set indicating which entries of $\widetilde{\textbf Y}_t$ are observed.
We exploit such a low-rank property and formulate the observation matrix complement problem as an LRMC problem:
\begin{equation}\label{LRMC}
\begin{aligned}
 \underset{\widehat{\textbf{Y}}_t}{\mathop{\text{min}}}\quad\text{rank}(\widehat{\textbf{Y}}_t)\text{ } \quad \text{s.t.}~\Big\vert\Big\vert{{\mathcal P}_{\bm\Omega }}(\widehat{\textbf{Y}}_t-\widetilde{\textbf{Y}}_t)\Big\vert\Big\vert^2_F<\varepsilon, 
\end{aligned}
\end{equation}
where $\Vert\cdot\Vert_F$ represents the Frobenius norm; $\widetilde{\textbf{Y}}_t$ is the incomplete data matrix. ${\mathcal P}_{\bm\Omega}$ is the associated sampling operator, which projects onto the corresponding subspace, ${\bm\Omega \in \mathbb R^{\mathcal L_1 \times \mathcal L_2}}$ represents the sampling domain, which is a binary indexing matrix. Specifically, the low-rank matrix $\widehat{\textbf Y}^{\star}$ can be recovered  via a nuclear-norm minimization:
\begin{equation}\label{Y-LRMC} {\widehat{\textbf Y}^\star } \triangleq \arg \min _{\widehat{\textbf Y} } \quad \left\Vert \mathcal P_{\bm\Omega }(\widehat{\textbf Y}) -\mathcal P_{\bm\Omega }({\widetilde{\textbf Y}})\right\Vert _{F}^{2} + \lambda \left\Vert \widehat{\textbf Y} \right\Vert _{\ast }, \end{equation}
where $\lambda$ is a positive constant that can be obtained by singular value decomposition (SVD). After recovering $\widehat{\textbf Y}_t^{\star}$, the refined $\widehat{\textbf H}_t$ can be estimated as
\begin{equation}\label{H_estimation}
\begin{aligned}
\widehat{\textbf H}_t=({\textbf W}_t^{\dagger})^{-1}\widehat{\textbf Y}_t^{\star}({\textbf F}_t)^{-1}.
\end{aligned}
\end{equation}
\indent
As in (\ref{Y-LRMC}), before the observation matrix ${\widetilde{\textbf Y}}_t$ is stably completed, it is prominent that only a rough channel estimation $\widetilde{\textbf H}_t=({\textbf W}_t^{\dagger})^{-1}\widetilde{\textbf Y}_t({\textbf F}_t)^{-1}$ can be obtained.  Next, we will solve the channel estimation problem under incomplete observations in two consecutive phases.
\section{Two-Phase Channel Estimator }
\begin{figure}[t]
	\centering 
	\includegraphics[width=6.5cm]{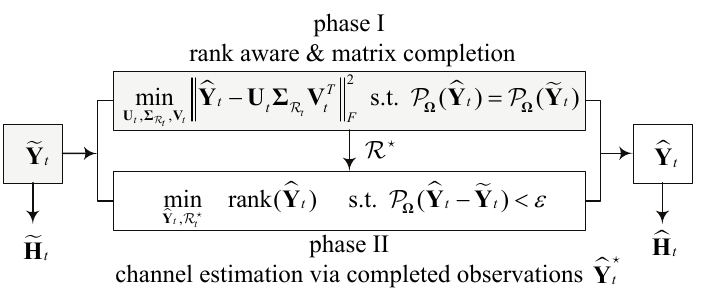}\\
	\caption{The proposed two-phase channel estimation framework.}
\end{figure}
\subsection{Phase I: Dynamic Rank Estimation and Matrix Completion}
\textit{1) Dynamic Rank Estimation}: To estimate the rank of ${\widetilde {\bf Y}}_t$, we follow the approach proposed in [10], given as
\begin{equation}
\label{rank-esti}
\begin{aligned}
\mathop {\min }\limits_{{\bf{U}_t},{{\bm{\Lambda }}_{{\widetilde{\mathcal R}_t}}},{\bf{V}_t}} \Big\| {{{\widehat {\bf{Y}}}_t} - \underbrace{{\bf{U}}_t{{\bf{\Lambda }}_{{\widetilde{\mathcal R}_t}}}{{\bf{V}}^T _t}}_{{\widetilde {\textbf{Y}}_t}}} \Big\|_F^2\quad{\rm{ s}}{\rm{.t}}{\rm{.}}\;\;{{\cal P}_{\bm{\Omega }}}({\widehat {\bf{Y}}_t}) = {{\cal P}_{\bm{\Omega }}}({\widetilde {\bf{Y}}_t}),
\end{aligned}
\end{equation}
where the factor matrices ${\textbf U}_t\in {\mathbb C}^{\mathcal L_1\times {\widetilde{\mathcal R}}_t}$ and ${\textbf V}_t\in {\mathbb C}^{ {\widetilde{\mathcal R}}_t \times \mathcal L_2}$ have orthogonal columns; ${\bm{\Lambda }_t}\in {\mathbb C}^{ {\widetilde{\mathcal R}}_t\times  {\widetilde{\mathcal R}}_t}$ is a diagonal matrix computes the best rank-$\widetilde{\mathcal R}_t$ approximation of the trimmed matrix via sparse SVD. $\widehat{\mathcal R}_t$ is estimated as the singular value index if the ratio between two consecutive singular values is minimum. 
\\
\indent
Generally, ${\widetilde {\bf{Y}}_t}$ can be further represented as the weighted sum of $\widehat{\mathcal R}_t$ factorized rank-one matrices
\begin{equation}
\label{rank-svd}
\begin{aligned}
{\widetilde {\bf{Y}}}_t={\bf{U}}_t\text{diag}(\bm\lambda){{\bf{V}}}^T _t=\sum_{\imath=1}^{\widehat{\mathcal R}_t}\lambda_{\imath}\cdot\textbf u_{\imath}\textbf v_{\imath},
\end{aligned}
\end{equation}
where the singular values $\lambda_{\imath}$ have been sorted ($\lambda_1 \le\lambda_2 \le···\lambda_{\widehat{\mathcal R}_t} \le 0$), and $\textbf u_{\imath}$ and $\textbf v_\imath$ denote the $\imath$th left and right singular vectors. Consider SVD is a special rank-one approximation whose factors. The weight vector $\bm\lambda$ derived from (14) requires rank-aware refinement to enforce low-rank constraints. 
\\
\indent
Following (\ref{rank-esti}), the effective rank size of ${\widetilde{\textbf Y}}_t$ is estimated as ${\widehat{\mathcal R}}_t$. Let $\{\lambda_i\}^{\widehat{\mathcal R}_t}_{\imath=1}$ denote the singular values sorted in descending order; the effective rank is determined by
\begin{equation}
\widehat{\mathcal R}_t = \max\left\{ k \in \mathbb{Z}^+ : \sum_{{\imath}=1}^k \lambda_{\imath} \geq \xi \sum_{{\imath}=1}^n \lambda_{\imath} \right\},
\label{eq:rank_est}
\end{equation}
where $\xi \in (0,1)$ controls the energy retention ratio. The resultant low-rank approximation guarantees
\begin{equation}
\|{\textbf Y}_t - \widetilde{\textbf{Y}}_t\|_F \leq \sqrt{1 - \alpha^2} \|\textbf{Y}\|_F,
\label{eq:error_bound}
\end{equation}
where $\alpha\in [0,1]$ is model error factor. The estimated rank subsequently governs the sparsity level in orthogonal matching pursuit and measurement matrix dimension $\bm{\Phi}_t \in \mathbb{C}^{({\mathcal N}_{\text{MS}}\times {\mathcal N}_{\text{BS}}) \times \widehat{\mathcal R}_t}$.
\\
\indent
To handle abrupt rank changes caused by user mobility, we further integrate an online rank predictor into R1MC. The rank sequence $\widehat{\mathcal R}_t$ is modeled as a Markov process, approximated with a discrete-time AR process of order $\jmath$ as:
\begin{equation} \label{eq:AR_model}
\widehat{\mathcal R}_t = \sum\limits_{\jmath=1}^{\jmath} a_{\jmath} \widehat{\mathcal R}_{t-\jmath} + b_0 Z_t,
\end{equation}
where $a_{\jmath}$ is $\jmath$th weighting factor; $b_0$ is a scaling factor; $Z_t\sim\mathcal N(0,1)$ and all $Z_t$'s are independent and identically distributed (i.i.d.). 
To efficiently tackle the dynamic rank estimation issue, the actuarial estimates of $\widehat{\mathcal R}_t$ are based on the following general procedure: The term autoregressive stems from the fact that $\widehat{\mathcal R}_t$ is predicted from the $\jmath$ previous $\widehat{\mathcal R}_t$’s through a regression equation. As a beginning, it may be assumed that the initial value $\widehat{\mathcal R}_t$ could be a fixed rank $\widehat{\mathcal R}_{t-1}$. The typical strategy is to choose $\widetilde{\mathcal R}_{t-1}$ for the initial matrix recovery process.
\\
\indent
\textit{2) Matrix Completion}: Once the rank estimation is completed, we can start with matrix completion for the incomplete matrix ${\mathcal P}_{\bm\Omega}$ with the index matrix set $\bm\Omega$. Recall (\ref{rank-svd}), this is somewhat a specific low-rank matrix decomposition for matrix completion, a.k.a rank-one matrix completion (R1MC). Let $\vert\vert{\bf u}_{\imath}\vert\vert_2\triangleq1$ and $\vert\vert{\bf v}_{\imath}\vert\vert_2\triangleq1$ for ${\imath}=1, ..., \widehat{\mathcal R}_t$, so that ${\bf u}_{\imath}$ and ${\bf u}_{\imath}$ form an orthonormal basis for all of $\mathbb R^n$, respectively. By applying a generalized least absolute shrinkage and selection operator (Lasso), or an $\ell_1$-norm regularization on $\bm\lambda$, the Lasso-type matrix recovery problem can be stated as follows:
\begin{align}\label{R1_admm}&\min _{\bm {\lambda},\{\textbf u_{{\imath}},\textbf v_{{\imath}} \}_{{\imath}=1}^{{\widehat{\mathcal R}}_t}} ~  \frac {1}{2} \Big \|{ \widehat{\text {Y}}_t - \sum _{\imath=1}^{{\widehat{\mathcal R}}_t} \lambda_{{\imath}} \textbf {u}_{{\imath}} \textbf {v}_{{\imath}}^T }\Big \|_{F}^{2}\notag +\mu \| \bm {\lambda} \|_{1}\\&\quad ~~~\text {s.t.} \qquad~\mathcal {P}_{\bm {\Omega }} (\widehat{\textbf {Y}}_t) = \mathcal {P}_{\bm {\Omega }} (\widetilde{\textbf {Y}}_t),  \end{align}
where $\mu$ is an augmented Lagrangian parameter, $\bm{\lambda}=({{\lambda}_{1}},...,{{\lambda}_{\widehat{\mathcal R}_t}})$.
\\
\indent
The above optimization problem for the multi-dimensional variables is preferred to use the alternating direction method of multipliers (ADMM)\cite{b45,Boyd04}. For simplification, we denote ${\mathcal B}_q=\{\lambda_q,{\textbf{v}}_q,{\textbf{u}}_q\}$, ${q=1,...,\widehat{\mathcal R}_t}$ as the $q$th block coordinate descent variants. Further, with the augmented Lagrangian function, \eqref{R1_admm} is converted into an unconstrained minimization problem, such that the corresponding algorithm takes the form
\begin{equation}\label{R2}
\begin{aligned}
{\mathscr {L}}(\textbf{Y},\bm{\lambda},{\textbf{v}}_{\imath},{\textbf{u}}_{\imath};\textbf{M})&=\frac{1}{2}\Big\vert\Big\vert \widehat{\textbf{Y}}_t-\underset{\imath=1}{\overset{\widehat{\mathcal R}_t}{\mathop{\sum }}}\,{{\lambda}_{{\imath}}}\cdot{\textbf{v}_{{\imath}}}\textbf{u}_{{\imath}}^{{T}}\Big\vert\Big\vert _{F}^{2}+\\ 
&\text{tr}(\textbf{M}^\dagger(\widehat{\textbf{Y}}_t-\underset{\imath=1}{\overset{\widehat{\mathcal R}_t}{\mathop{\sum }}}\,{{\lambda}_{i}}\cdot{\textbf{v}_{{\imath}}}\textbf{u}_{{\imath}}^{{T}}))+\mu\parallel \bm{\lambda}{{\parallel }_{1}}, 
\end{aligned}
\end{equation}
where $\textbf{M}\in {{\mathbb{R}}^{\mathcal L_1\times \mathcal L_2}}$ is the Lagrange multiplier, $\mu>0$ is a penalty parameter. The standard ADMM iteration should first fix ${{\textbf{Y}}}$ and $\textbf{M}$, and then sequentially update $\left\{ {{\textbf{v}}_{{\imath}}},{{\textbf{u}}_{{\imath}}} \right\}_{\imath=1}^{\widehat{\mathcal R}_t}$ and $\bm{\lambda}$. Correspondingly, \eqref{R2} can be represented as
\begin{equation}\label{R3}
\mathscr{L}(:,\mathcal B_q;:)=\frac{1}{2}\left\vert\left\vert {\textbf{Y}}_q-{{{\lambda}}_{q}}\cdot{{\textbf{v}}_{q}}\textbf{u}_{q}^{{T}}\right\vert\right\vert_{F}^{2}+\mu\vert \lambda_q \vert,
\end{equation}
where $\textbf{Y}_q=\widehat{\textbf{Y}}_t-\underset{{\imath}=1}{\overset{q-1}{\mathop{\sum }}}\,{{{\lambda}}_{{\imath}}}\cdot{{\textbf{v}}_{{\imath}}}\textbf{u}_{{\imath}}^{{T}}$ is the residual of the approximation. Using the block coordinate descent (BCD) method, where $\left\{ {{\textbf{v}}_{{\imath}}},{{\textbf{u}}_{{\imath}}} \right\}_{{\imath}=1}^{\widehat{\mathcal R}_t}$ and $\bm{\lambda}$ are divided into $\widehat{\mathcal R}_t$ blocks $\{\{{{\textbf{v}}_{1}},{{\textbf{u}}_{1}},{{{\lambda}}_{1}}\},...,\{{{\textbf{v}}_{\widehat{\mathcal R}_t}},{{\textbf{u}}_{\widehat{\mathcal R}_t}},{{{\lambda}}_{\widehat{\mathcal R}_t}}\}\}$, ${\mathcal{B}}_q^{(k+1)}$ can be updated by
\begin{equation}
{\mathcal{B}}_q^{(k+1)}=\left\{{shrinkage_{\mu}(a)},\frac{{\textbf{Y}_q^T}^{(k)}{\textbf u}_q^{(k+1)}}{\lambda_q},\frac{\textbf{Y}_q^{(k)}{\textbf v}_q^{(k)}}{\lambda_q}\right\},
\end{equation}
where $a= \big \langle \textbf {Y}_{q}^{k}, ~{\textbf {v}_ {q}}^{(k+1)} {\textbf {u}_{q}^T }^{(k+1)} \big \rangle$, ${shrinkage}_{\mu }  (\cdot)$ is a soft shrinkage operator, which is defined as $sign(a)\max(\vert a \vert-\mu,0)$.
With the help of ADMM iteration, it allows us to deal with matrix completion problems in a unified manner, i.e., solving
\begin{equation}
\widehat{\textbf{Y}}_t^{\star}=\arg\min \vert\vert \widehat{\textbf{Y}}_t\vert\vert_*\quad\text{s.t.}\;\,\vert\vert {\mathcal P}_{\bm{\Omega}}(\textbf{Y}_t)-{\mathcal P}_{\bm{\Omega}}(\widehat{\textbf{Y}}_t)\vert\vert_F\le \varepsilon.
\end{equation}
\\
\indent
Note that when $\mathcal B_q$ is obtained by (21), the estimated rank will constrain the subsequent channel estimation and design of the measurement matrix.
\subsection{Phase II: Channel Estimation via Completed Observations}
In Phase II, our main goal is to accurately estimate the parameters of ${\textbf H}_t$. Without loss of generality, we denote the parameter set as ${\mathcal C}=\{{\bm\theta},\textbf g,\bm\tau\}$ and further refine the recovery results by directly minimizing the reconstruction errors.  The estimation procedure studied here can be briefly indicated as follows. When the observations are incomplete, the corresponding estimates remain coarse. Once the rank is obtained, we remove the $\ell_1$-norm regularization, and refine the estimates. By leveraging the reconstructed \(\widehat{\textbf{Y}}_t\) and rank $\widehat{\mathcal R}_t$, we estimate the virtual beam-space channel \(\bm{\Phi}_t,\widehat{\textbf{Y}}_t\) by exploiting the inherent sparsity of \(\textbf{H}_t\). 
 \\
 \indent
 Formally, the channel gain matrix can then be estimated by solving the following sparse recovery problem:
 \begin{equation}\label{H_estimation}
\begin{aligned}
{\bm{\mathcal G}}_{\mathcal C}^{\star}=\arg \underset{\bm{\mathcal G}_{\mathcal C}}{\mathop{\min }}\,||\widehat{\textbf{H}}_{t}-\textbf{A}_{\text{BS}}^{\dagger}\otimes\textbf{A}_{\text{MS}}\bm{\mathcal G}_{\mathcal C}|{{|}^{2}},\text{s.t.}~||\bm{\mathcal G}_{\mathcal C}|{{|}_{0}}\leq \widehat{\mathcal R}_t^2,
 \end{aligned}
\end{equation}
 The sparsity pattern of $\bm{\mathcal G}^{\star}_{\mathcal C}$ also reveals the correspondence of the set of $\mathcal C$, i.e., AoAs, AoDs and delay. Finally, once $\bm{\mathcal G}^{\star}$ is estimated, the low-rank channel matrix is reconstructed as
 \begin{equation}\label{H_estimation}
\begin{aligned}
\widehat{\textbf H}_t=({\textbf A}_{\text {MS}}^{\dagger})^{-1}\bm{\mathcal G}^{\star}\widehat{\textbf Y}_t^{\star}{\textbf A}_{\text{BS}}. 
\end{aligned}
\end{equation}
To quickly recover the non-zero-valued entries of the channel matrix, we adopt the classical batch OMP algorithm for low-latency sparse channel recovery. The complexity of the OMP algorithm primarily depends on the number of iterations and the rate of residual error updates. In each iteration, for each column $\textbf g$ in the dictionary, we compute the residual projection from the previous iteration onto the column. We then minimize the difference between the actual signal $\vert\vert\widehat{\textbf H}_{t}^{(k)}-{\bm\Phi}_t^{(k)}\widehat{\textbf Y}_t^{\star}\vert\vert$ and the approximate signal derived from the current solution $\widehat{\textbf H}_{t}^{(k)}$. The current estimate is obtained via least squares estimation on the subdictionary formed by the selected atoms, and the residuals are updated accordingly. In this step, we identify the solution that minimizes $\vert\vert\widehat{\textbf H}_{t}^{(k)}-{\bm\Phi}_t^{(k)}\widehat{\textbf Y}_t^{\star}\vert\vert$ over the support set $\Gamma^{(k+1)}\leftarrow\{{\mathcal B}_q^{(k)},{\bm\Phi}_t^{(k)}\}$ which becomes the next candidate solution vector. Finally, the residual vector is updated to $\widehat{\textbf H}_{t}^{(k+1)}$. 
\section{Numerical Results}
The proposed rank-aware channel estimation framework is evaluated through comprehensive simulations. 
The system operates at a 28 GHz carrier frequency with 15 kHz subcarrier spacing and 0.1 $\mu$s sampling interval. Channel dynamics are generated by the open-source MATLAB-based NYU Channel Model Simulator (NYUSIM) [12]. The performance is evaluated via the normalized mean squared error (NMSE), defined as $\text{NMSE}=\text E\left[\frac{\Vert {\textbf H}_t-\widehat{\textbf H}_t \Vert}{\Vert  {\textbf H}_t\Vert_F}\right]$.
\\
\indent
Figs. 3 and 4 compare the successful recovery probability versus SNR. In Fig. 3, the proposed method achieves superior recovery probability at the high-SNR regime, outperforming the SPC-TDSC scheme proposed in \cite{b43}, which only exploits delay-domain sparsity. The results validate the advantage of joint spatio-temporal correlation exploitation over pure delay-domain sparsity approaches. As shown in Fig. 4, the proposed method achieves a 3.8 dB NMSE reduction compared to SOMP and SPC-TDSC. This improvement stems from the proposed method's exploitation of temporal channel correlation, in contrast to quasi-static channel assumptions, demonstrating scalable performance across antenna configurations.
\\
\indent
%
Ablation studies in Fig. 5 validate the critical role of rank constraints. Disabling rank feedback ("Unranked-FOMP") degrades NMSE by 2 dB at $\nu=0.1$, while fixed-rank assumptions increase BER. These results emphasize the necessity of dynamic rank adaptation in high-mobility environments.
Finally, Fig. 6 demonstrates bit error rate (BER) superiority, where the proposed scheme achieves better BER performance than CNN-based methods. In contrast, SOMP fails to attain BER even at 25 dB SNR, highlighting its practical limitations. 
\begin{figure}[htbp]
	\centering 
	\includegraphics[width=7.0cm]{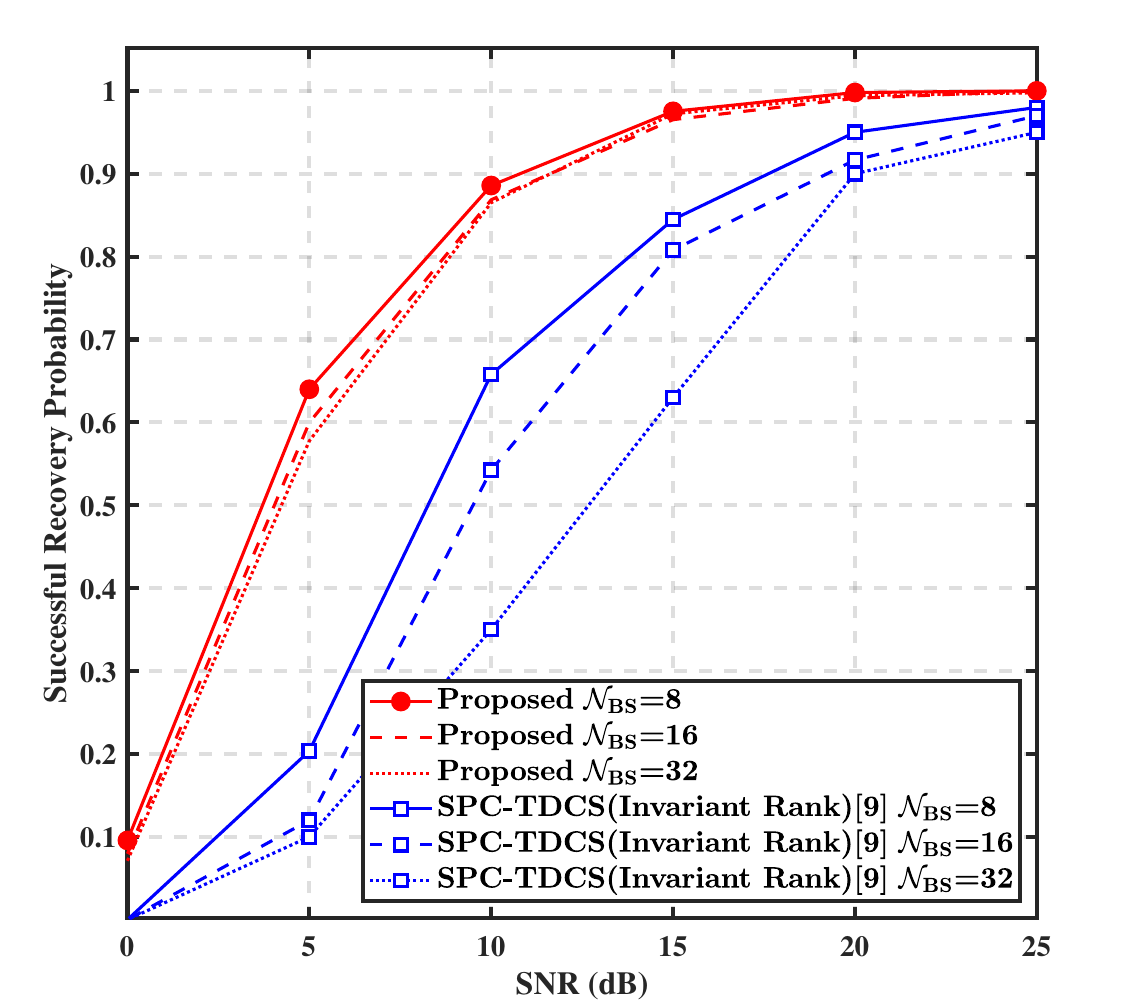}
	\caption{Recovery probability vs SNR, $v=120$ km/h, ${{\mathcal N}_\text{BS}}=8$, ${{\mathcal N}_\text{MS}}=8$.}

\end{figure}
\begin{figure}[htbp]
	\centering 
	\includegraphics[width=7.0cm]{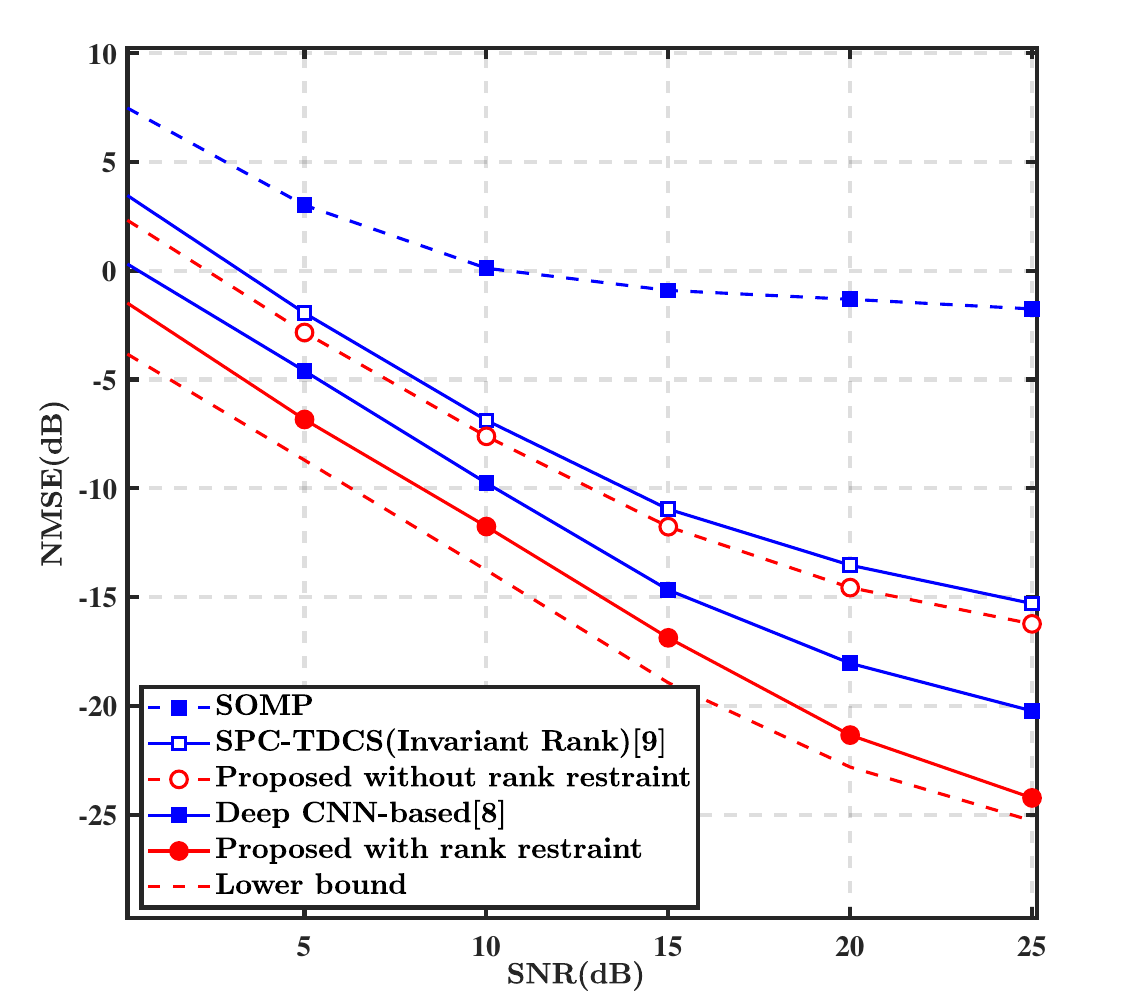}
	\caption{NMSE vs. SNR, $v=120$ km/h, ${\mathcal N}_\text{MS}=8$, ${{\mathcal N}_\text{BS}}=8$ and ${\mathcal N}_\text{BS}=64$.}

\end{figure}
\begin{figure}[htbp]
	\centering 
	\includegraphics[width=7.0cm]{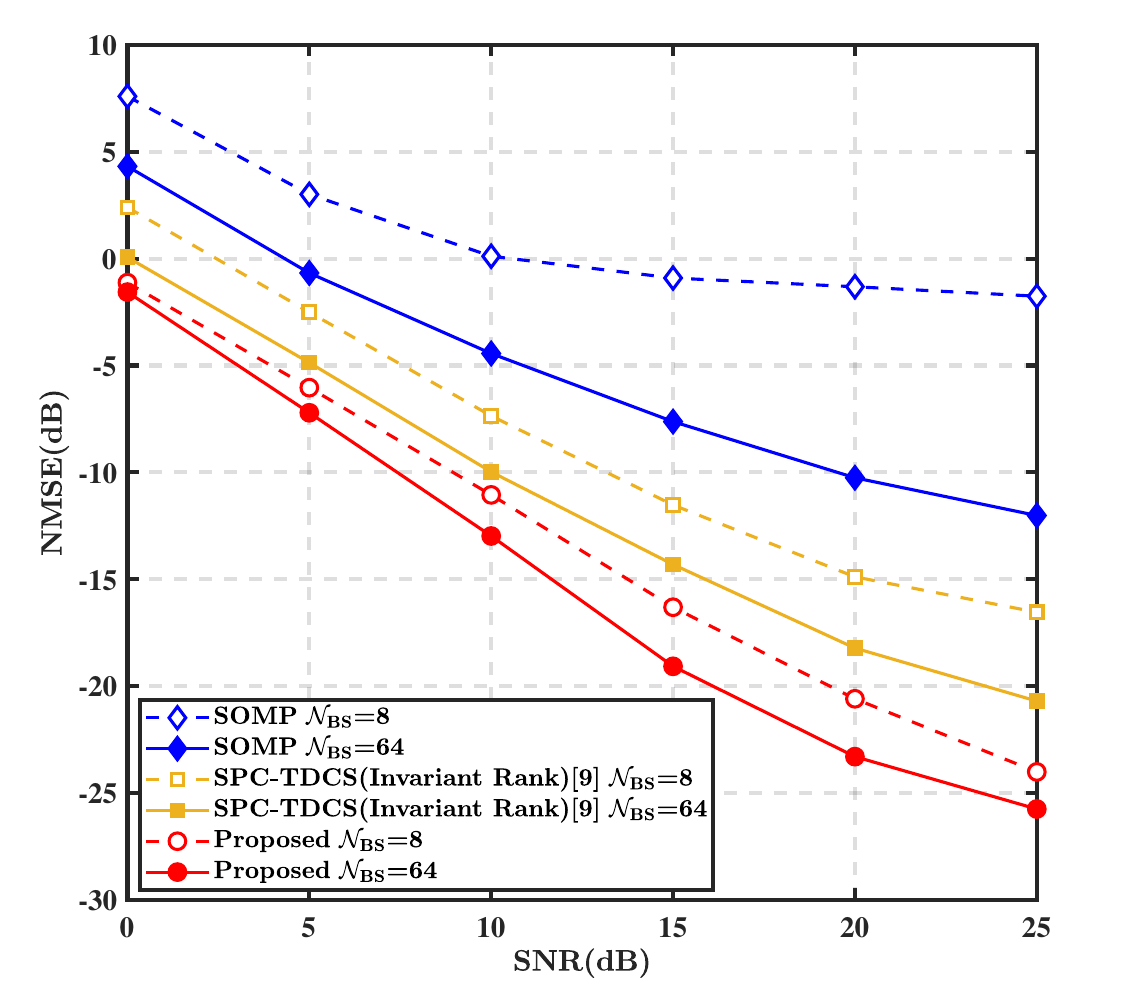}
	\caption{NMSE vs. SNR, $v=120$ km/h, ${{\mathcal N}_\text{BS}}=8$, ${{\mathcal N}_\text{MS}}=8$.}

\end{figure}
\begin{figure}[htbp]
	\centering 
	\includegraphics[width=7.0cm]{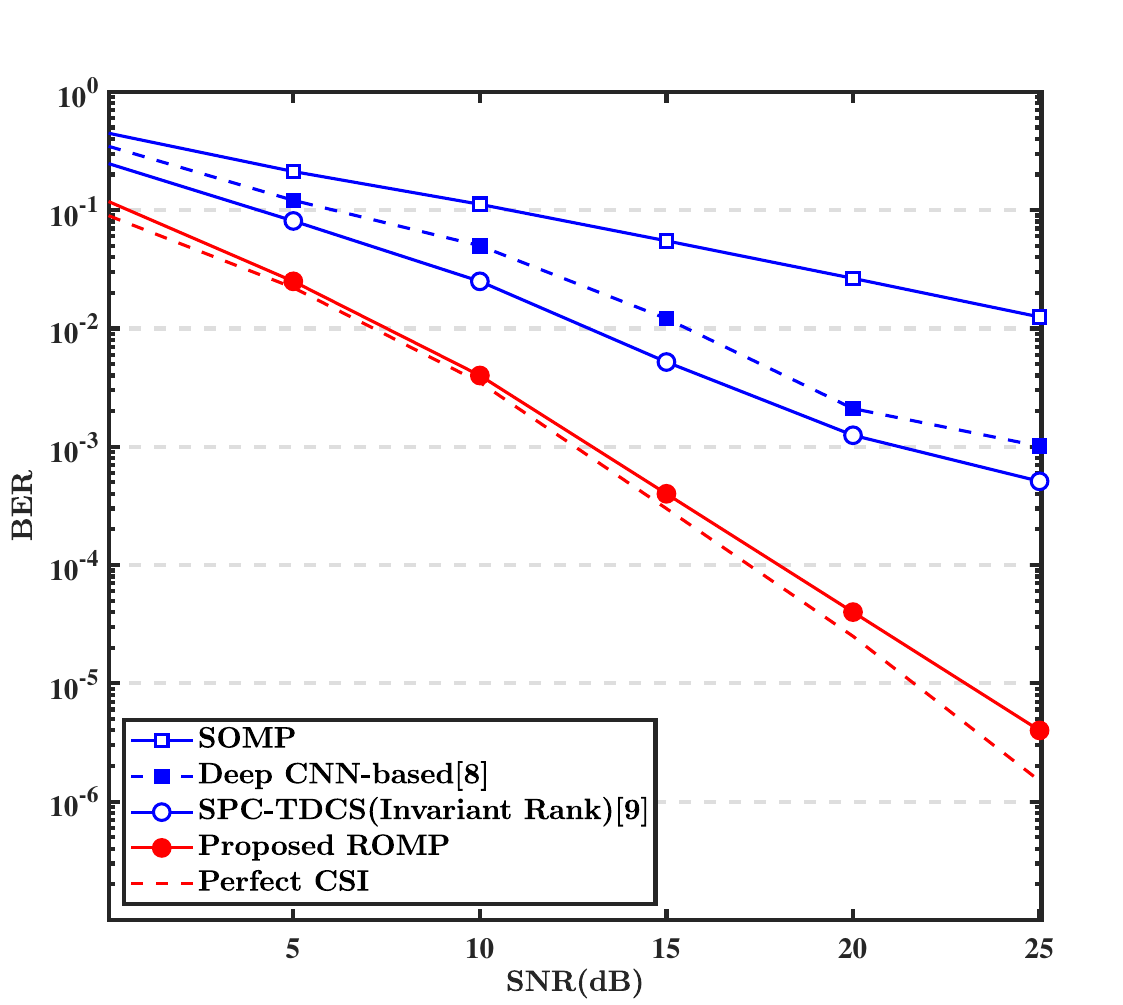}
	\caption{BER vs. SNR, $v=120$ km/h, ${{\mathcal N}_\text{BS}}=8$, ${{\mathcal N}_\text{MS}}=8$.}
	
\end{figure}
\section{Conclusion}
%
This paper presents a novel rank-aware channel estimation framework for mmWave MIMO systems operating in time-varying channels. The method decouples the estimation into two phases, observation matrix completion and sparse recovery, specifically designed for mmWave channel estimation challenges. Theoretical analysis proves that our framework reduces measurement requirements compared to conventional convex relaxation-based compressed sensing methods, which solely exploit channel sparsity. Extensive simulations validate the theoretical claims and demonstrate the framework's superiority in estimation accuracy and computational efficiency over state-of-the-art benchmarks.

\end{document}